# «СЛАБЫЕ» ИЗМЕРЕНИЯ И СВЕРХСВЕТОВАЯ КОММУНИКАЦИЯ

А.В. Белинский[1], А.К. Жуковский[2]

*Московский государственный университет имени М. В. Ломоносова, физический факультет, кафедра математического моделирования и информатики, кафедра физики Земли. Россия, 119991, Москва, Ленинские горы, д. 1, стр. 2.*
*E-mail: belinsky@inbox.ru[1], andrez@rambler.ru[2]*

Предложен вариант эксперимента с коррелированной парой частиц в запутанном (entangled) состоянии, который показывает, что, в случае слабых и/или невозмущающих измерений одной из частиц, можно осуществить передачу информации со скоростью, не ограниченной световой.

Слабые измерения (см., напр., [1] и цитируемую там литературу), в отличие от обычных квантовых измерений, обладают тем характерным свойством, что они позволяют получать частичную информацию о квантовом состоянии объекта без редукции вектора состояния, т.е. после измерений последний остается без изменения. В [1], например, таким образом, удалось оценить энергию фазового кубита. Но к чему приведет возможность такого измерения в случае коррелированной пары частиц в запутанном (entangled) состоянии, произведенного над одной из них? Постараемся пофантазировать и оценить возможные последствия такой перспективы, хотя в [2-4] и утверждается о невозможности измерения одиночной квантовой частицы без последующей редукции всей квантовой системы. Это, конечно, правильно, но редукция бывает разной, и можно попробовать сделать ее действительно слабой.

Коррелированную пару фотонов легко получить при помощи параметрического рассеяния света. При этом коррелированы могут быть как состояния поляризации, спины, так и суммарная энергия обеих частиц, хотя каждая из них будет иметь определенный разброс энергий в зависимости от толщины кристалла, в котором происходит параметрический процесс, и, соответственно, ширины спектра генерируемого излучения.

Итак, предположим, что нам удалось измерить слабым образом, скажем, энергию первого фотона (см. рис. 1). В случае измерения состояния поляризации рассуждения будут аналогичны. Если эти слабые измерения производить периодически, или лучше – непрерывно, то пока второй фотон не измерен сильным образом, т.е. с редукцией вектора состояния обеих частиц, слабые измерения первого будут давать полный разброс измеряемой наблюдаемой величины во всем возможном диапазоне. Как только мы произведем сильное измерение над второй частицей, вектор состояние обеих частиц нелокальным образом моментально редуцирует (см., напр., недавний эксперимент [5], фактически подтверждающий более ранние [6,7]), и первая частица примет определенное значение измеряемой величины без всякого разброса.

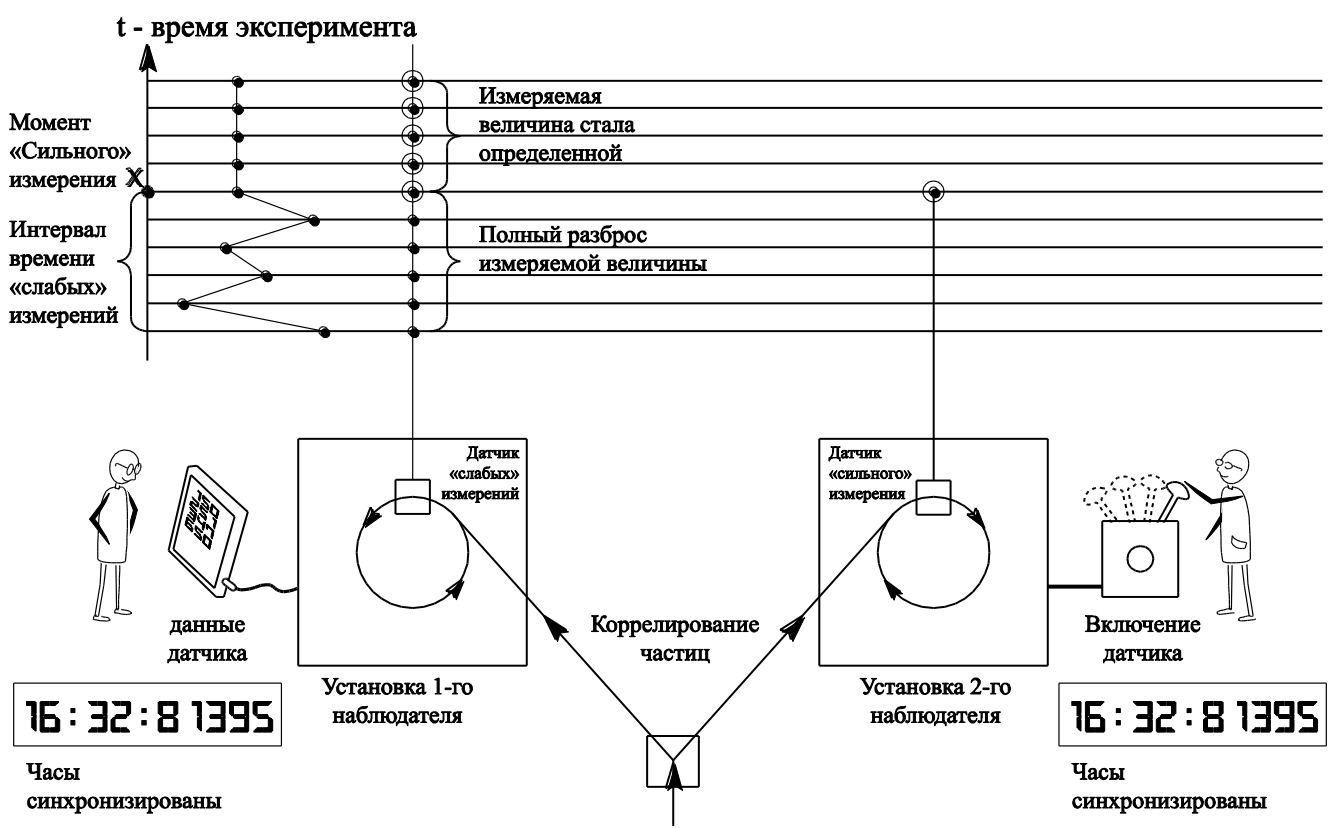

Рис. 1. Схема эксперимента по передаче момента времени измерения со сверхсветовой скоростью. Слабые измерения фотона в установке 1-го наблюдателя будут давать полный разброс измеряемой наблюдаемой величины во всем возможном диапазоне до тех пор, пока 2-й наблюдатель не произведет «сильное» измерение коррелированной частицы. После этого первый наблюдатель, через 2-3 такта измерений (после того, как измеряемая величина станет определенной), зафиксирует момент времени данного измерения.

При этом следует учесть, что сами слабые измерения дают не точное значение измеряемой величины. Поэтому, их разброс должен быть, разумеется, меньше исходного разброса возможных значений энергии или состояний поляризации частиц. Но если это условие выполнено, то наблюдатель первой частицы может моментально, без ограничений на световую скорость, узнать момент времени сильного измерения, произведенного наблюдателем второй частицы (см., также [8]).

Итак, если, скажем в условленный времени момент *t* (а часы у обоих наблюдателей должны быть синхронизированы) второй наблюдатель произведет сильное измерение, кардинально редуцирующее вектор состояния пары наших частиц, то мы припишем этому событию значение 1, если нет, то 0. Таким образом, один бит информации от одного наблюдателя к другому будет передан со сверхсветовой скоростью. Не нужно говорить, что последующими событиями такого рода можно передавать информацию на неограниченные расстояния, при этом находясь в режиме абсолютной конфиденциальности, поскольку, если по пути следования сигнала пропадет хоть один фотон, то это будет означать возможность утечки информации и последующей необходимости сменить канал связи. Однако, в случае шифрования сообщения специальным кодом, возможность «похищения» информации в случае «пропажи» небольшой части фотонов будет все равно практически невозможна.

Приведенные рассуждения неоспоримо свидетельствуют о том, что любое предложение по реализации измерения квантового состояния одиночной квантовой частицы без редукции квантовой системы в целом немедленно приведет к возможности сверхсветовой коммуникации. А это противоречит соответствующей теореме о невозможности передачи информации со сверхсветовой скоростью при помощи пары квантовых частиц в запутанном состоянии [9]. С другой стороны, полностью исключить возможность таких коммуникаций все же, как представляется, преждевременно [10].

Рассмотрим, например, пару запутанных фотонов, коррелированных по поляризации. Их вектор состояния равен

$$|\psi\rangle = \frac{1}{\sqrt{2}}\left(|1\rangle_x^a |1\rangle_x^b |0\rangle_y^a |0\rangle_y^b + |0\rangle_x^a |0\rangle_x^b |1\rangle_y^a |1\rangle_y^b\right). \tag{1}$$

Здесь индексы «*a*» и «*b*» относятся, соответственно, к первому и второму фотону запутанной пары, а взаимно ортогональные поперечные направления *x* и *y* определяют ортогональные направления поляризации. Структура этого вектора состояния такова, что, хотя направление поляризации *x* и *y* каждого из фотонов пары «*a*» или «*b*» равновероятны, между собой они строго коррелированы, поскольку их плоскости поляризации всегда совпадают при регистрации. Такие состояния обычно приготавливают с помощью параметрического рассеяния света (см., напр., [11] и цитируемую там литературу).

Направим каждый из фотонов пары на призму Волластона, разделяющую взаимно ортогональные поляризации на два отдельных канала. Фактически она работает как светоделитель, а для фотонов с абсолютно случайной поляризацией – как 50%-ный светоделитель. Покажем в начале, что при таком разделении на два канала фотон до момента его регистрации находится сразу в обоих каналах. Рассмотрим для этого интерферометр Маха-Цендера, на вход которого подается одиночный фотон в фоковском состоянии (рис. 2). Уберем в начале, второй светоделитель, расположенный перед фотодетекторами. Включим детекторы, которые начнут регистрировать одиночные фотоотсчеты либо в одном, либо в другом канале с равной вероятностью 1/2.

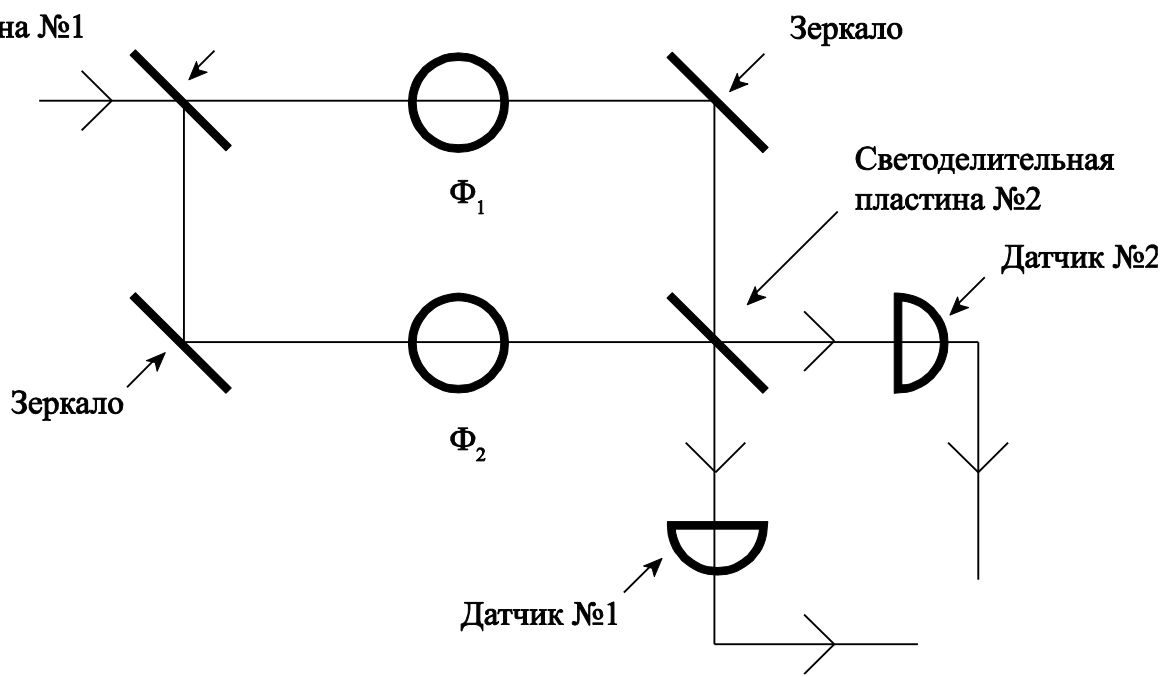

Рис. 2. Схема интерферометра Маха-Цендера. Вероятность фотоотсчетов на детекторах описывается гармонической функцией $P_{1,2} = \frac{1}{2}[1 \pm \cos(\Phi_1 - \Phi_2)]$, где $\Phi_1$ и $\Phi_2$ — фазовые задержки в плечах интерферометра, а знак в этом выражении зависит от того, какой детектор осуществляет регистрацию.

Что произойдет после того, как мы вернем второй светоделитель на место? Вероятность фотоотсчетов на детекторах теперь будет описываться гармонической функцией (см., напр., [12]):

$$P_{1,2} = \frac{1}{2}[1 \pm \cos(\Phi_1 - \Phi_2)], \tag{2}$$

где $\Phi_1$ и $\Phi_2$ — фазовые задержки в плечах интерферометра. Знак в этом выражении зависит от того, какой детектор осуществляет регистрацию. Эту гармоническую функцию нельзя представить в виде суммы вероятностей

$$P_{1,2}(\Phi_1, \Phi_2) \neq P(\Phi_1) + P(\Phi_2). \tag{3}$$

Следовательно, после первого светоделителя фотон будет присутствовать в обоих плечах интерферометра одновременно, хотя в первом акте эксперимента он регистрировался только в одном. Это связано с тем, что редукция вектора состояния квантовой системы происходит только в момент регистрации фотона, а до этого он находится в обоих каналах.

Тот же результат можно доказать и другим способом. При разности фаз $0+2\pi m$ или $\pi+2\pi m$ одиночные фотоны будут появляться только на одном из детекторов. Вероятность появления фотонов на втором будет нулевой. Перекроем одно из плеч интерферометра. Фотоотсчеты, у до того «безмолвного» фотодетектора, появятся. Это означает, что до перекрытия поле присутствовало в этом плече в каждой реализации, иначе вероятность появления фотоотсчетов на рассматриваемом фотодетекторе не была бы нулевой. Аналогично доказывается присутствие поля в каждой реализации и в другом плече интерферометра. Следовательно, результат измерения месторасположения фотона (в каком плече находится фотон?) при убранном втором светоделителе до момента измерения не определен, поскольку фотон находится в обоих плечах сразу (см., также [12]).

Есть еще одно строгое доказательство пребывания фотона в обоих плечах интерферометра одновременно. Поместим вместо или в качестве фазовых задержек две идентичные нелинейные среды, обладающие кубичной нелинейностью, в которых происходит фазовая самомодуляция (ФСМ), т.е. изменение показателя преломления сред под действием находящегося в них света. Это могут быть, например, кварцевые волокна (рис. 3). Фотон, проходя через них, должен приобретать дополнительный фазовый набег, который неизбежно скажется на результате интерференции.

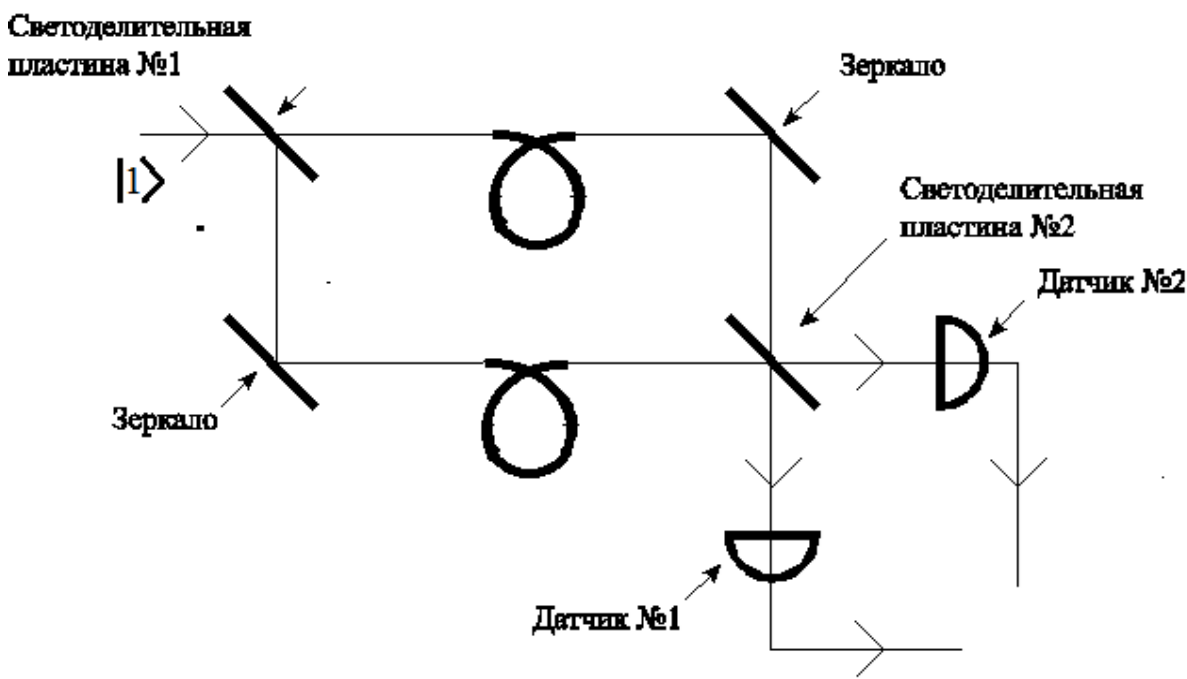

Рис. 3. Схема интерферометра Маха-Цендера с идентичными нелинейными волокнами в каналах.

Пусть в отсутствие излучения фазовые набеги в плечах были одинаковы. Тогда, посылая в интерферометр единичный фотон, мы имеем две альтернативы: либо фотон пройдет только одно плечо, и разность фаз изменится за счет нелинейного набега фазы в этом плече, либо фотон пройдет оба плеча, нелинейные фазовые набеги в которых будут одинаковы, так что разность фаз не изменится. В последнем случае мы увидим появление фотона лишь на одном из выходов интерферометра.

Входную монохроматическую моду в фоковском состоянии $|1\rangle$ будем описывать оператором уничтожения фотона $\hat{a}_1$, а вакуумную моду $|0\rangle$ на втором входе – оператором

$\hat{a}_0$. После первого 50%-ного светоделителя также рассматриваем две моды, описываемые операторами $\hat{a}_2$, $\hat{a}_3$ в представлении Гейзенберга:

$$\hat{a}_2 = \frac{\hat{a}_1 + \hat{a}_0}{\sqrt{2}}, \qquad \hat{a}_3 = \frac{\hat{a}_1 - \hat{a}_0}{\sqrt{2}}. \qquad (4)$$

Далее учитываем действие керровской нелинейности. Устойчивое поперечное распределение интенсивности в кварцевых волокнах можно рассматривать как моду излучения, а сам четырехфотонный процесс описывать одномодовым гамильтонианом (см., напр., [13] и цитируемую там литературу):

$$\hat{H} = \frac{\hbar}{2}\chi^{(3)}\hat{a}^+\hat{a}^+\hat{a}\hat{a}, \qquad (5)$$

где $\chi^{(3)}$ – коэффициент кубичной нелинейности, пронормированный по числу фотонов. Нелинейный отклик полагаем мгновенным.

Соответствующий оператор эволюции квантового состояния в представлении Шредингера равен

$$\hat{U} = \hat{I}\exp\left(-i\frac{\overline{\chi}}{2}\hat{a}^+\hat{a}^+\hat{a}\hat{a}\right) = \hat{I}\exp\left(-i\frac{\overline{\chi}}{2}\hat{n}(\hat{n}-1)\right), \qquad (6)$$

где $\overline{\chi} = \chi^{(3)}t$, а время эволюции $t$ связано с длиной волокна $l=vt$, $v$ – скорость распространения моды в волокне, $\hat{n}(t)$ – оператор числа фотонов.

В представлении Гейзенберга оператор уничтожения фотона моды поля подчиняется уравнению $i\hbar\frac{d\hat{a}}{dt} = [\hat{a}, \hat{H}]$, откуда $\hat{a}(t) = e^{-i\overline{\chi}\hat{a}^+(0)\hat{a}(0)}\hat{a}(0)$, а в нашем случае

$$\hat{a}'_2 = e^{-i\overline{\chi}\hat{a}_2^+\hat{a}_2}\hat{a}_2, \qquad \hat{a}'_3 = e^{-i\overline{\chi}\hat{a}_3^+\hat{a}_3}\hat{a}_3. \qquad (7)$$

Соответственно, две выходные моды интерферометра:

$$\hat{a}'_0 = \frac{\hat{a}'_2 - \hat{a}'_3}{\sqrt{2}}, \qquad \hat{a}'_1 = \frac{\hat{a}'_2 + \hat{a}'_3}{\sqrt{2}}. \qquad (8)$$

Найдем средние значения чисел фотонов на выходах интерферометра:

$$\langle \hat{n}_0 \rangle \equiv \langle \hat{a}'^+_0 \hat{a}'_0 \rangle = 0, \qquad \langle \hat{n}_1 \rangle \equiv \langle \hat{a}'^+_1 \hat{a}'_1 \rangle = 1. \qquad (9)$$

Итак, мы наблюдаем интерференцию с нулевой разностью фаз, значит, фотон прибывает в обоих каналах одновременно.

Если же фотон поступает с другого входного канала (сверху на рис. 3), то и на выходе он окажется в другом канале с вероятностью единица, следовательно, сам факт его интерференции с нулевой+2π*m* или π+2π*m* разностью фаз, а, значит, и пребывания одновременно в обоих плечах интерферометра останется неизменным. Таким образом, и для линейной суперпозиции пребывания фотона на обоих входах вывод остается тем же, а это и есть как раз случай прохождения фотоном с абсолютно случайной поляризацией призмы Волластона.

Перейдем теперь к невозмущающему измерению первого фотона. Установим в оба выходных канала после призмы Волластона среды с кубичной нелинейностью (рис. 4), в которых происходит фазовая самомодуляция. Поскольку оператор $\hat{n}(t)$ при ФСМ является инвариантом во времени, величина числа фотонов при ФСМ является невозмущаемой наблюдаемой и может быть невозмущающим образом измерена. Как это сделать? Подадим на входы нелинейных сред с кубичной нелинейностью (тех же кварцевых волокон, напр.) помимо измеряемых сигналов "$a_1, a_2$" еще и слабые пробные моды "$p_1, p_2$" равной средней интенсивности, по измерению разности фаз которых попытаемся определить: находится ли первый фотон "$a$" в состоянии суперпозиции до сильного измерения второго фотона "$b$", или же в одном из каналов после редукции вследствие такого сильного измерения. Действительно, поскольку сигналы "$a_1, a_2$" изменяют оптическую плотность и показатель преломления нелинейных сред, такие изменения плотности можно «прощупать» обычным интерферометром, например, Маха–Цендера. Этим интерферометром можно измерить косинус и синус разности фаз в плечах в зависимости от того, какие постоянные фазовые задержки в них установлены (имеются в виду линейные фазовые задержки, не зависящие от интенсивности света). Если постоянная разность фаз нулевая+$2\pi m$, то будет измерен косинус, а если $\pi/2+2\pi m$ – синус. Для этого детекторы, установленные на выходах интерферометра, должны быть включены в разностную схему, дающую разностный фототок (см., напр., [13]).

Итак, если фотон "$a$" находится в состоянии суперпозиции и присутствует в обоих каналах одновременно, то фазовые задержки в нелинейных средах одинаковы, как и в рассмотренном выше нелинейном интерферометре Маха-Цендера на рис. 3. Если же редукция уже произошла, то фотон "$a$" окажется лишь в одном канале, и разность фаз будет определяться фазовым набегом именно в этом канале. Но можно ли таким образом выяснить, в каком канале находится фотон "$a$"? Измеряя синус разности фаз – можно, поскольку синус «чувствует» знак аргумента, по которому однозначно определяется место расположения фотона. Однако, произведя такое измерение, мы радикально редуцируем вектор состояния (1) раньше, чем сильное измерение, которое будет произведено над фотоном "$b$". В итоге, такое измерение не окажется слабым и все испортит.

Совсем другое дело, если мы будет регистрировать косинус разности фаз. В этом случае информация о том в каком плече находится фотон, будет «стерта», аналогично «квантовому ластику» (см., напр., [14]) за счет четности косинуса. Может быть эта аналогия с «квантовым ластиком» и не совсем удачна, но такое измерение не редуцирует квантового состояния (1) радикально, оставляя свободу фотону находиться в обоих каналах сразу. В то же время, система почувствует момент редукции за счет сильного измерения фотона "$b$", поскольку изменится разность фаз в каналах, которую для получения наибольшего контраста можно сделать равной $\pi$ за счет соответствующего выбора длины нелинейных сред, хотя это и не обязательно, если чувствительность измерений и так достаточна.

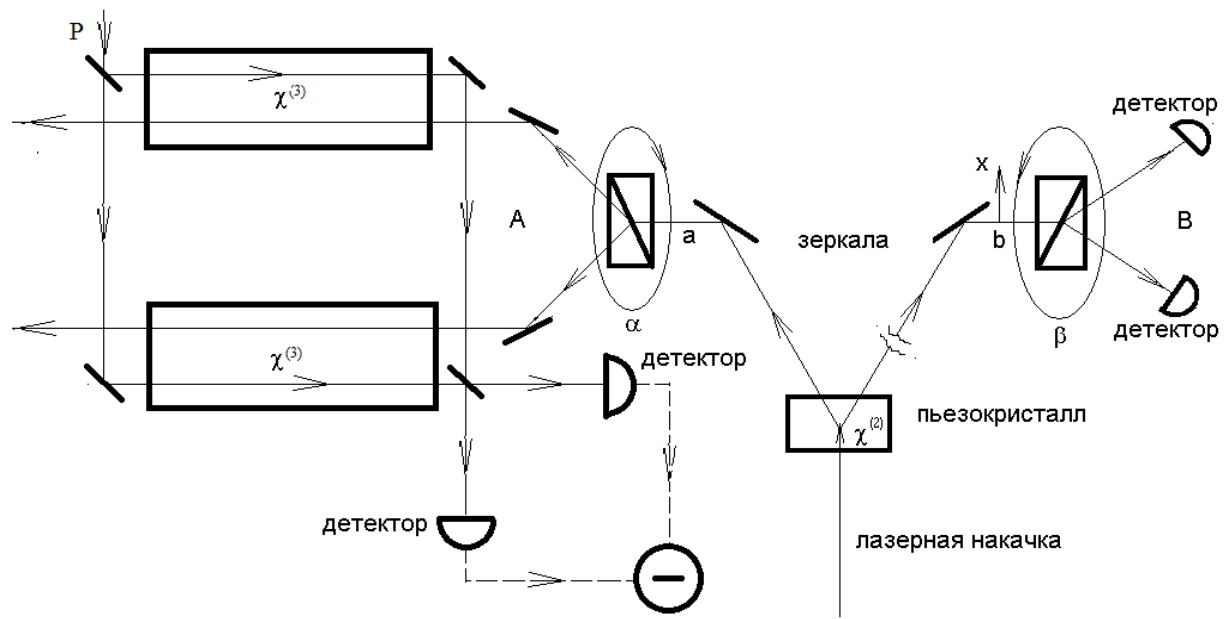

Рис. 4. Схема измерения наблюдателем А момента редукции вектора состояния в результате сильного измерения наблюдателем В.

Итак, если приведенные соображения верны, то используя идею эксперимента, схематично представленного на рис.1 в его конкретной реализации, иллюстрируемой рис. 4, мы сможем передать один бит информации от наблюдателя В к наблюдателю А практически мгновенно.

# «WEAK» MEASUREMENTS AND SUPRALUMINAL COMMUNICATION

## A.V. Belinsky[1], A.K. Zhukovsky[2]

*Department of computer modeling and informatics, Department of Physics of the Earth Faculty of Physics, Lomonosov Moscow State University. Moscow 119991, Russia*
*E-mail: belinsky@inbox.ru[1], andrez@rambler.ru[2]*

There is suggested a version of the experiment with a correlated pair of particles in the entangled state. The experiment demonstrates that, in the case of weak and/or non-demolition measurements of one of the particles, it is possible to transmit information with a speed not limited by velocity of light.

*Key words:* weak measurement, entanglement, nonlocality, reduction of the state vector, speed of light, supraluminal speed, phase self-modulation.

Сведения об авторах:

1. Белинский Александр Витальевич – доктор физико-математических наук, ведущий научный сотрудник, профессор; тел.: (495) 939-41-78,
   e - mail: belinsky@inbox.ru

2. Жуковский Андрей Кузьмич – физик, тел.: (985) 991-44-40, e-mail: andrez@rambler.ru